\documentclass[aps,prl,preprint,noshowpacs,superscriptaddress]{revtex4-1}
\usepackage{amsmath}
\usepackage{amssymb}
\usepackage{graphicx}
\usepackage[normalem]{ulem} 
\usepackage{color}

\begin{document}

\title{Oscillatory Non-collinear Magnetism Induced by Interfacial Charge Transfer in Metallic Oxide Superlattices}

\author{J. Hoffman}
\email{jhoffman@anl.gov}
\affiliation{Materials Science Division, Argonne National Laboratory, Argonne, Illinois 60439}

\author{B. J. Kirby}
\affiliation{NIST Center for Neutron Research, National Institute of Standards and Technology, Gaithersburg, Maryland 20899}

\author{J. Kwon}
\affiliation{Department of Materials Science and Engineering, University of Illinois at Urbana-Champaign, Urbana, Illinois 61801}

\author{J. W. Freeland}
\affiliation{Advanced Photon Source, Argonne National Laboratory, Argonne, Illinois 60439}

\author{I. Martin}
\affiliation{Materials Science Division, Argonne National Laboratory, Argonne, Illinois 60439}

\author{O. G. Heinonen}
\affiliation{Materials Science Division, Argonne National Laboratory, Argonne, Illinois 60439}

\author{P. Steadman}
\affiliation{Diamond Light Source, Diamond House, Harwell Science and Innovation Campus, Didcot, Oxfordshire, OX11 0DE, United Kingdom}

\author{H. Zhou}
\affiliation{Advanced Photon Source, Argonne National Laboratory, Argonne, Illinois 60439}

\author{C. M. Schlep{\"u}tz}
\affiliation{Advanced Photon Source, Argonne National Laboratory, Argonne, Illinois 60439}

\author{S. G. E te Velthuis}
\affiliation{Materials Science Division, Argonne National Laboratory, Argonne, Illinois 60439}

\author{J.-M. Zuo}
\affiliation{Department of Materials Science and Engineering, University of Illinois at Urbana-Champaign, Urbana, Illinois 61801}

\author{A. Bhattacharya}
\email{anand@anl.gov}
\affiliation{Materials Science Division, Argonne National Laboratory, Argonne, Illinois 60439}
\affiliation{Nanoscience and Technology Division, Argonne National Laboratory, Argonne, Illinois 60439}

\date{\today}

\begin{abstract}
Interfaces between correlated complex oxides are promising avenues to realize new forms of magnetism that arise as a result of charge transfer, proximity effects and locally broken symmetries. We report upon the discovery of a non-collinear magnetic structure in superlattices of the ferromagnetic metallic oxide La$_{2/3}$Sr$_{1/3}$MnO$_3$ (LSMO) and the correlated metal LaNiO$_3$ (LNO). The exchange interaction between LSMO layers is mediated by the intervening LNO, such that the angle between the magnetization of neighboring LSMO layers varies in an oscillatory manner with the thickness of the LNO layer.  The magnetic field, temperature, and spacer thickness dependence of the non-collinear structure are inconsistent with the bilinear and biquadratic interactions that are used to model the magnetic structure in conventional metallic multilayers.  A model that couples the LSMO layers to a helical spin state within the LNO fits the observed behavior.  We propose that the spin-helix results from the interaction between a spatially varying spin susceptibility within the LNO and interfacial charge transfer that creates localized Ni$^{2+}$ states. This provides a new approach to engineering non-collinear spin textures in metallic oxide heterostructures that can be exploited in devices based on both spin and charge transport. 
\end{abstract}



\maketitle
Oxide interfaces have attracted considerable interest in recent years, as the reconstruction of charge, orbital, and spin states on the nanometer scale gives rise to novel phenomena that range from interfacial superconductivity to multiferroic behavior.\cite{HIK12}  In this context, interfaces between \emph{metallic} oxides are particularly intriguing, as the large electronic compressibility, the relatively large bare dielectric constant, and band misalignment can work in concert to create significant interfacial charge transfer over a region of several unit cells.\cite{JG03, ZSJ05, BT10}  In metallic oxides derived from correlated Mott insulators, this effect can manifest latent electronic and magnetic instabilities, leading to new collective states near the interface.

While a large body of work has emerged on heterostructures that incorporate insulating complex oxides,\cite{SHM99, OMG02a, OH04, HSH07, LOV11, BMB11, GZS12, LCS13, FSH13, HTN13, CLS14} those created exclusively with metallic oxide constituents have been far less explored,\cite{NBK99, NDK00, KND01} despite the technological importance and wide range of behaviors observed in multilayers of conventional metals.  The discovery of giant magnetoresistance (GMR) \cite{BBF88, BGS89} and the subsequent demonstration of a tunable collinear exchange coupling in such structures,\cite{PMR90} opened new pathways to high-density magnetic data storage.  Multilayers with \emph{non-collinear} magnetic ordering, however, are rarer, as such structures require a delicate balance between exchange energies, which only occurs under special circumstances.\cite{S93a, S95, D98, S99b}  Engineering such non-collinear magnetic states at oxide interfaces presents new opportunities to explore novel effects, such as spin-transfer torque,\cite{BBK06} long-range superconducting proximity effects,\cite{BVE05} multiferroicity,\cite{CM07, TS10, TSN14} magnetic skyrmion phases,\cite{NT13,BRE14} and new phenomena that emerge from correlated electronic states not found in conventional metals.

In this work, we show that charge transfer at the interface between two correlated metallic oxides La$_{2/3}$Sr$_{1/3}$MnO$_3$ (LSMO) and LaNiO$_3$ (LNO) can stabilize a novel non-collinear magnetic structure.  In the bulk, LSMO is a ferromagnetic half-metal at low temperatures.  LNO is a correlated paramagnetic metal, where epitaxial strain, dimensional confinement, and interfacial exchange interactions are known to stabilize long-range charge and magnetic ordering in thin films and heterostructures.\cite{SML10, SGG11, CRL11, LOV11, BMB11, GZS12, FSH13, KWN14, KDN14}  Using polarized neutron reflectometry, we find that an intrinsically non-collinear magnetic structure develops in superlattices of LSMO and LNO, grown with oxide molecular beam epitaxy (MBE). The magnitude of non-collinearity (the angle between the magnetization of adjacent LSMO layers) oscillates with LNO thickness, without ever becoming antiferromagnetic (i.e., an angle of 180$^{\circ}$).  The magnetic field, temperature, and LNO-layer thickness dependence of the exchange coupling between the LSMO layers is incompatible with a model based on the combination of bilinear and biquadratic coupling that is widely used to characterize non-collinear magnetic interactions in conventional metallic heterostructures.  Rather, we show that the observed behavior is consistent with the development of a proper-screw-type magnetic order within LNO with a period of 5--7 unit cells along the (001) propagation direction.  This structure persists to near ambient temperatures, above the magnetic ordering temperature known for any of the rare-earth nickelates.  The helical spin structure proposed here is believed to result from a coupling between a momentum-dependent spin susceptibility $\chi(q)$ within the LNO layers and localized Ni$^{2+}$ spins produced by charge transfer at the LNO/LSMO interface, which we measure using x-ray spectroscopy.

Epitaxial superlattices of LSMO and LNO were coherently grown on (001)-oriented SrTiO$_3$ (STO) and (LaAlO$_3$)$_{0.3}$-(Sr$_2$TaAlO$_6$)$_{0.7}$ (LSAT) substrates at 600$^{\circ}$C using oxide MBE.  The number of repeats, which each consist of nine unit cells of LSMO ($c$ = 0.387 nm) and $n$ unit cells of LNO ($c$ = 0.382 nm), where $1 \leq n \leq 9$, was adjusted to achieve a total thickness of $\sim$60 nm ($\sim$160 unit cells).  The growth was monitored \emph{in situ} by reflection high-energy electron diffraction (RHEED), and the maxima in the oscillating specular spot intensity correspond to the deposition of a single unit cell layer.  This behavior was observed throughout the growth of the superlattice, indicating a layer-by-layer growth mode.  

\begin{figure}[htbp]
  \includegraphics[width=\columnwidth]{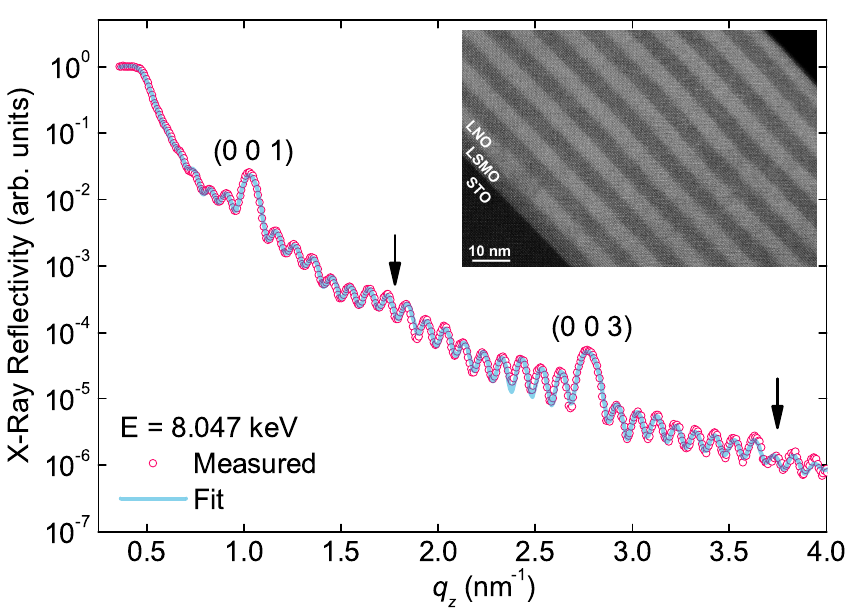}
	\caption{\textbf{Structural characterization of (LaNiO$_3$)$_n$/(La$_{2/3}$Sr$_{1/3}$MnO$_3$)$_9$ superlattices.} Measured x-ray reflectivity and fit for the $n = 9$ superlattice on SrTiO$_3$.  The arrows mark the position of the even-order superlattice reflections, which are strongly suppressed by the structural symmetry of the sample. \textbf{(Inset)} High-resolution transmission electron micrograph of the same superlattice.}
	\label{fig:fig_xrd}
\end{figure}

The structure of the superlattices was characterized by x-ray reflectivity (XRR), high-resolution x-ray diffraction (XRD), and $Z$-contrast scanning transmission electron microscopy (STEM) measurements.  Figure \ref{fig:fig_xrd} shows XRR results for the $n$ = 9 superlattice, which are representative of all the samples studied here.  From XRR data, we determine that the thickness of each superlattice period is within 0.5\% of the designed value and that the LNO/LSMO interfacial roughness is less than one unit cell.  Additional details about the growth and characterization may be found in the online supplementary material.

\begin{figure}[htbp]
  \includegraphics[width=\columnwidth]{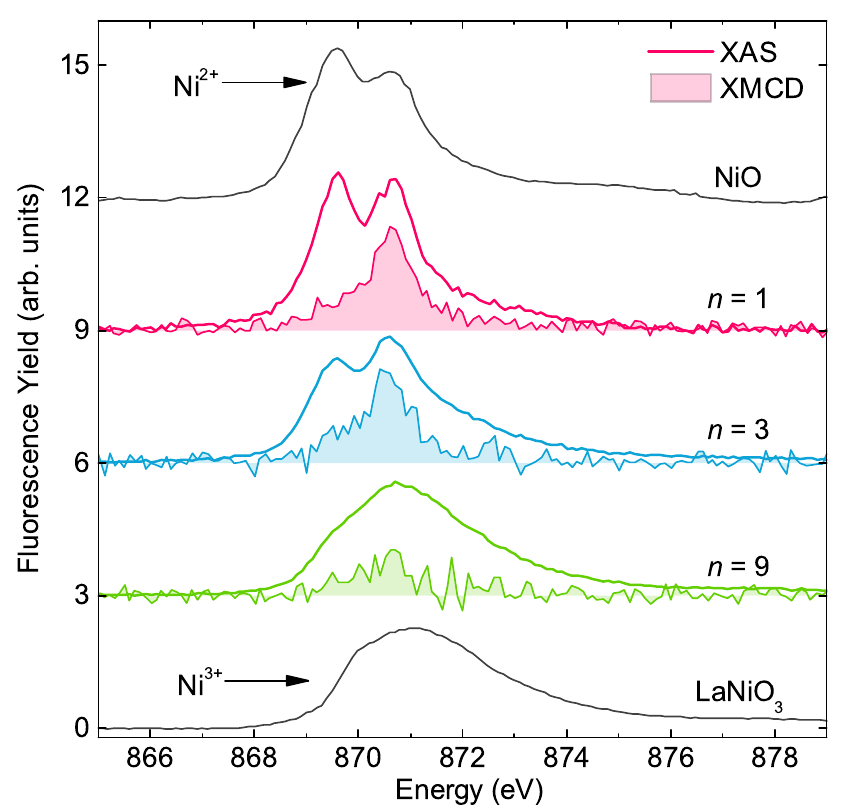}
	\caption{\textbf{Interfacial charge transfer and magnetic dichroism.} X-ray absorption spectroscopy at the Ni $L_2$-edge for (LaNiO$_3$)$_n$/(La$_{2/3}$Sr$_{1/3}$MnO$_3$)$_9$ superlattices (thick lines) showing progression from Ni$^{3+}$ to Ni$^{2+}$ as $n$ is varied from 9 to 1.  The shaded regions show the x-ray magnetic dichroism over the same energy range, confirming the existence of magnetism on the Ni sites.}
	\label{fig:fig_XAS}
\end{figure}

To probe the electronic structure and magnetism within the LNO spacer layer, we performed x-ray absorption spectroscopy (XAS) and x-ray magnetic circular dichroism (XMCD) measurements around the Ni $L_2$-edge. Figure \ref{fig:fig_XAS} shows the results for a series of LNO/LSMO superlattices with different LNO layer thicknesses, as well as NiO (Ni$^{2+}$) and LaNiO$_3$ (Ni$^{3+}$) reference spectra.   The evolution of the Ni $L_2$ peak shape and position shows an unambiguous change in Ni valence state from nearly Ni$^{3+}$ when $n$ = 9 to predominantly Ni$^{2+}$ for $n$ = 1.  This result is consistent with charge-transfer confined to a few unit cells at the interface, in agreement with previous studies on manganite/nickelate interfaces.\cite{SNG12,HTN13}  Complementary measurements at the Mn $L_{2,3}$-edge are compatible with a mixture of Mn$^{3+}$ and Mn$^{4+}$  valence states, as expected for this composition of LSMO. While we do not observe any significant differences between the Mn XAS spectra for the samples measured here, signatures of a predominantly Mn$^{4+}$ valence state were observed in previous work on LaNiO$_3$/LaMnO$_3$ superlattices with 2 unit cell thick LaMnO$_3$ layers, consistent with a transfer of electrons from the manganite to the nickelate layer.\cite{HTN13}

The XMCD spectra measured in an applied in-plane field of $\sim$250 mT at 110 K (shown as shaded regions in Fig. \ref{fig:fig_XAS}) confirm that interfacial doping leads to a net magnetic moment within the LNO layer.  We note that the total fluorescence yield technique used here has a probe depth that is comparable to the sample thickness and that the XMCD spectra shown in Fig. \ref{fig:fig_XAS} are not normalized by the volume of Ni probed, which changes as $n$ is varied.  The magnetization of the LNO layers in the $n$ = 3 superlattice was investigated with x-ray resonant magnetic scattering (XRMS) measurements, which are consistent with a modulation of the Ni magnetization commensurate with the superlattice structure (Fig. S3).

\begin{figure*}[htbp]
  \includegraphics[width=\textwidth]{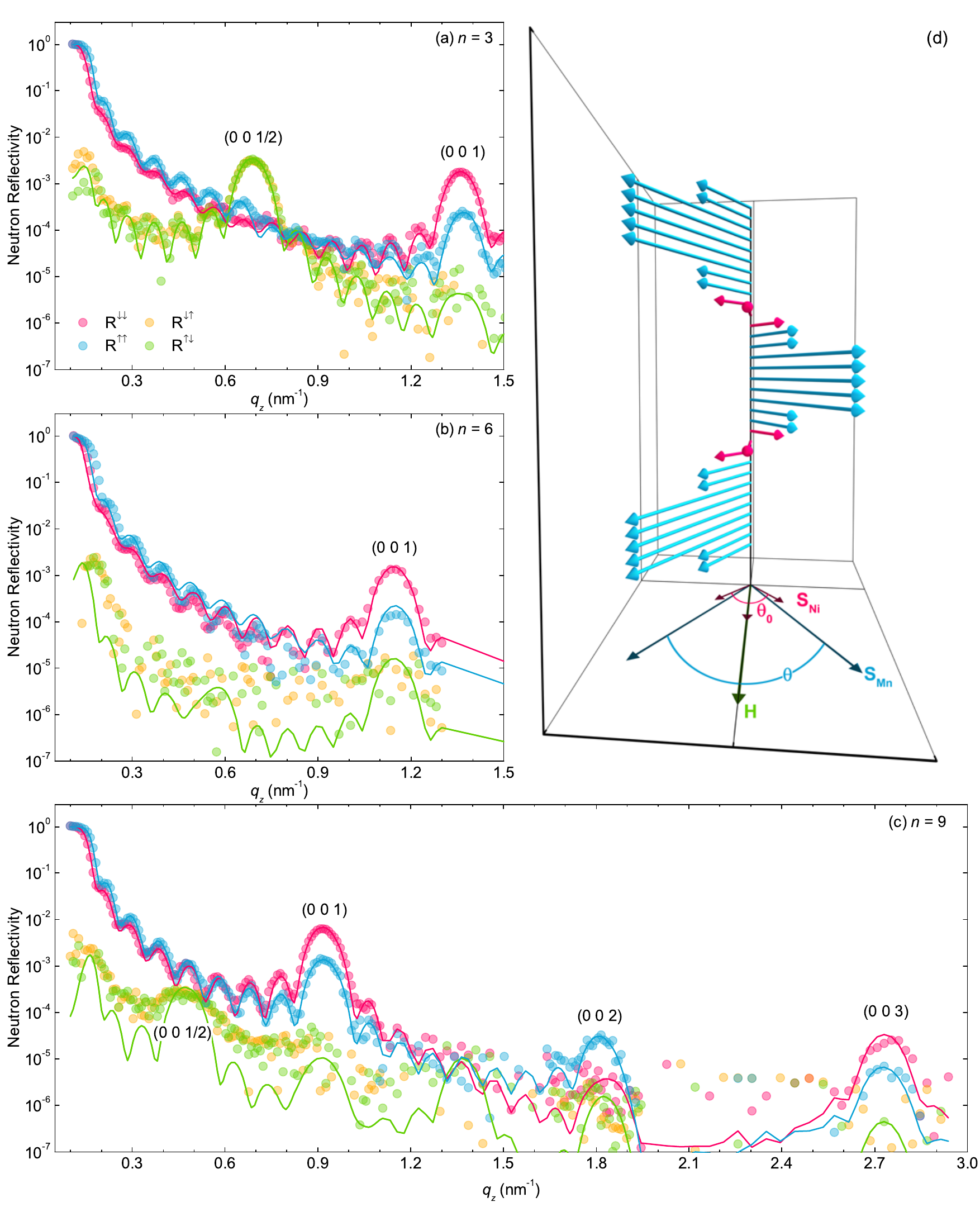}
	\caption{\textbf{Polarizared neutron reflectometry for (LaNiO$_3$)$_n$/(La$_{2/3}$Sr$_{1/3}$MnO$_3$)$_9$ superlattices on SrTiO$_3$.} Measured (symbols) and fit (lines) PNR spectra for superlattices with $n$ = 3 \textbf{(a)}, $n$ = 6 \textbf{(b)}, and $n$ = 9 \textbf{(c)}.  All spectra measured at $T$ = 110 K with an applied in-plane field of 1.2 mT.  \textbf{(d)} Schematic magnetic structure within the $n$ = 3 superlattice.  The LSMO and LNO layers are shown in blue and pink, respectively.}
	\label{fig:fig_pnr}
\end{figure*}

To further explore the magnetization profile of the superlattices, we carried out polarized neutron reflectometry (PNR) measurements using the polarized beam reflectometer at the Center for Neutron Research at the National Institute of Standards and Technology.   The samples were cooled from room temperature in a 5 mT in-plane field applied along the [100] direction, which is parallel to $\mathbf{P}$, the polarization axis of the incident neutrons.  Neutron scattering shows a three-fold increase in mosaic spread below the cubic to tetragonal structural transition in STO, which is known to occur at $\sim$105 K.  PNR measurements of the superlattices grown on STO were therefore carried out at temperatures above 110 K.  We are able to determine the depth profile of the magnitude and orientation of the in-plane magnetization within each layer by measuring both the non-spin-flip (NSF) reflectivities R$^{\uparrow\uparrow}$ and R$^{\downarrow\downarrow}$ and the spin-flip (SF) reflectivities R$^{\uparrow\downarrow}$ and R$^{\downarrow\uparrow}$.  Here, the superscripts denote the initial and final neutron spin states.  The NSF reflectivities depend on both $\rho(z)$, the nuclear scattering length density and $M\left(z\right)$, the in-plane magnetization.  The SF reflectivities are only sensitive to $M_{\perp}\left(z\right)$, the projection of the magnetization that is perpendicular to the polarization direction and parallel to the interfaces.

The PNR measurements reveal a strongly modulated magnetization within the superlattices.  Figure \ref{fig:fig_pnr}(a-c) shows the PNR spectra measured at $T$ = 110 K in an applied field of 1.2$\pm$0.5 mT as a function of $q_z$, the momentum transfer along the surface normal, for the superlattices with $n$ = 3, 6, and 9 grown on STO.  In all three samples, we observe splitting between the NSF reflectivities at the critical edge and at the superlattice Bragg reflections, indicating a modulated profile that is commensurate with the superlattice period.  Within the Born approximation, the NSF Bragg reflection splitting signals a modulation of the in-plane magnetization component along the field direction.

Due to the symmetry of the $n$ = 9 superlattice, the chemical contribution to the even-order Bragg peaks is strongly suppressed, as evidenced by the XRR measurements in Fig. \ref{fig:fig_xrd}, further confirming the high structural quality of the superlattices.  In the non-spin-flip PNR spectra shown in Fig. \ref{fig:fig_pnr}(c), however, we observe a pronounced (002) peak at $q_z$ $\sim$1.8 nm$^{-1}$, demonstrating that the magnetic profile does not follow the chemical structure exactly.  To account for this observation, we have considered three scenarios for the interface magnetization: \emph{i}) an induced magnetization within the LNO at the interface, \emph{ii}) a reduced magnetization within the interfacial LSMO, \emph{iii}) a combination of these two effects.  Through detailed fitting using \textsc{REFL1D},\cite{KKM12} we find that an induced magnetization on the interfacial Ni sites alone is not sufficient to quantitatively explain the observed (002) peak.  Rather, our fitting shows the magnetization of the LSMO is reduced from its bulk value of $\sim$550 kA/m ($\sim$3.5 $\mu_{\mathrm{B}}$/Mn) to $\sim$315 kA/m ($\sim$2.0 $\mu_{\mathrm{B}}$/Mn) within 1--2 unit cells of the interface, in agreement with the length scale for interfacial charge transfer that we found in our XAS measurements (Fig. \ref{fig:fig_XAS}).

Furthermore, we see clear evidence for \emph{non-collinear} alignment between the magnetization of adjacent LSMO layers.  In the superlattices with $n$ = 3 and $n$ = 9, our measurements reveal an additional peak in both SF reflectivities at $q_z$ $\sim$0.7 nm$^{-1}$ ($n$ = 3) and 0.5 nm$^{-1}$ ($n$ = 9), which correspond to the positions of the $(0 0 \tfrac{1}{2})$  reflections.  Thus, there exists a magnetization component that lies \emph{perpendicular} to the applied field direction, which is modulated with \emph{twice} the periodicity of the superlattice.  The $(0 0 \tfrac{1}{2})$ peak in the $n$ = 9 superlattice is weaker than in the $n$ = 3 superlattice, consistent with a smaller coupling angle in this sample.  In the superlattice with $n$ = 6 (Fig. \ref{fig:fig_pnr}(b)), however, the $(0 0 \tfrac{1}{2})$ peak is completely absent, indicating that adjacent LSMO layers are ferromagnetically aligned, even at the lowest measurement fields.  In the $n$ = 9 superlattice, the $(0 0 \tfrac{1}{2})$ peak is broadened relative to the (001) peak, which may indicate that $\theta$, the angle between the magnetization of adjacent LSMO layers, is not constant throughout the thickness of the superlattice.  We have attempted fitting of the PNR data for each superlattice with a number of model structures and find that $\theta$ is insensitive ($\pm 5^{\circ}$) to the details of the model used.  

\begin{figure}[htbp]
  \includegraphics[width=\columnwidth]{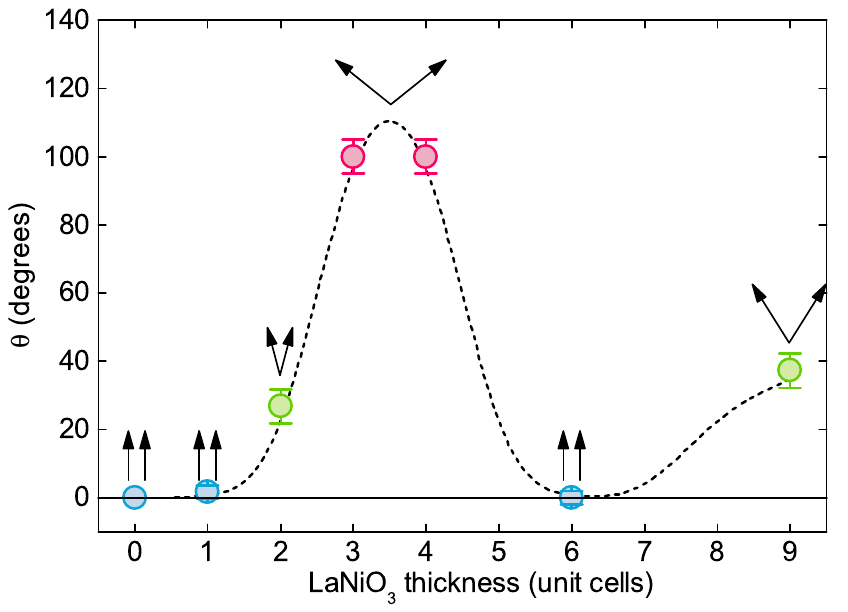}
	\caption{\textbf{Oscillatory non-collinear coupling in (LaNiO$_3$)$_n$/(La$_{2/3}$Sr$_{1/3}$MnO$_3$)$_9$ superlattices.}  Dependence of coupling angle, $\theta$, on LNO thickness for superlattices grown on SrTiO$_3$ at $T$ = 110 K and $H$ = 1.2 mT. Line is a guide to the eye.}
	\label{fig:fig_theta_vs_n}
\end{figure}

To study how the coupling angle varies with LNO thickness, we carried out PNR measurements on superlattices with $1 \leq n \leq 9$.  Figure \ref{fig:fig_theta_vs_n} shows how the value of $\theta$ obtained from fitting the PNR spectra shown in Fig. \ref{fig:fig_pnr} changes as the thickness of LNO is varied.  The coupling angle is found to oscillate with a period of $\sim$5--7 unit cells.  Remarkably, we find that the LSMO magnetization never attains complete antiferromagnetic (180$^{\circ}$) alignment, as previously reported for La$_{2/3}$Ba$_{1/3}$MnO$_3$/LaNiO$_3$ multilayers.\cite{NBK99, NDK00, KND01}  Rather, the non-collinearity reaches a maximum of around $100^{\circ}$ for $n$ = 3 and 4.  This behavior is qualitatively different from that found for conventional metallic multilayers, such as Fe/Cr, where non-collinear magnetic alignments are typically only observed in a narrow regime of spacer layer thickness, where the interlayer exchange coupling transitions between collinear ferromagnetic and antiferromagnetic alignments.\cite{RSH91}

To uncover the origin of the surprising behavior shown in Fig. \ref{fig:fig_theta_vs_n}, we examine two possible models for the interlayer exchange coupling.\footnote{Slonczewski has proposed a third mechanism - the proximity magnetism model - to explain non-collinear interlayer coupling in systems with antiferromagnetic spacer layers, such as Cr and Mn.\cite{S95}  Such a model does not explain the observed variation in exchange coupling with spacer layer thickness, which is never antiferromagnetic.  Furthermore, we have not found evidence for A-type or G-type antiferromagnetic ordering within the LNO layers in neutron diffraction experiments.}  We first consider a phenomenological description based on bilinear (collinear) and biquadratic (non-collinear) effects, which is widely used to model the exchange interactions within conventional metallic multilayers\cite{S95, D98, S99b}: 
\begin{equation}
F_{\mathrm{BLBQ}}(\theta) = -J_{\mathrm{BL}} \cos(\theta) - J_{\mathrm{BQ}} \cos^2(\theta).
\label{eqn:eqn_FBLBQ}
\end{equation}
The bilinear term $J_{\mathrm{BL}}$ derives from the topology of the Fermi surface of the LNO spacer layer and oscillates in sign with LNO thickness, favoring either parallel or antiparallel alignment of the LSMO layers.  $J_{\mathrm{BQ}}$ is usually attributed to defects such as interfacial roughness, steps, or magnetic impurities in the spacer layer, and favors a 90$^{\circ}$ alignment for $J_{\mathrm{BQ}} < 0$.  In principle, an additional term due to anisotropy may be present.  However, as described in the supplementary information, SQUID magnetometry measurements confirm that the in-plane magnetocrystalline anisotropy in our samples is relatively weak (($\lvert K t_{\mathrm{LSMO}}\rvert \lessapprox 10^{-6}$ J m$^{-2} \ll \lvert J_{\mathrm{BL}} \rvert \sim \lvert J_{\mathrm{BQ}} \rvert \approx 10^{-5}$--$10^{-4}$ J m$^{-2}$), and may be ignored.  Furthermore, neutron scattering measurements on the $n = 3$ superlattice grown on LSAT with the magnetic field applied along the $\langle 110 \rangle$ direction yields the same value of $\theta$ as for measurements with the field applied along $\langle 100 \rangle$.  In the absence of anisotropy, the value of $\theta$ at $H = 0$ is determined by the ratio of $J_{\mathrm{BL}}$ to $J_{\mathrm{BQ}}$, favoring non-collinear alignment for $\lvert J_{\mathrm{BQ}}\rvert > \lvert J_{\mathrm{BL}}\rvert/2$.

As an alternative mechanism to explain the observed non-collinear exchange coupling, we propose the formation of a spin-helix within the LNO layers.  Helimagnetic states should arise generally for large, localized spins and a magnetic susceptibility $\chi(q)$ with peaks that favor a spatially oscillating magnetic order.\cite{K10}  Transfer of electrons from the LSMO into the LNO layer creates interfacial Ni$^{2+}$ states (Fig. \ref{fig:fig_XAS}), which are expected to form localized moments with $S = 1$, within a region of $\sim$2 unit cells of each interface.  Furthermore, photoemission measurements and first principles calculations indicate that the Fermi surface of nickelates have a nested character.\cite{NDK00,ECT09}  As a result, calculations of $\chi(q)$ for these materials have peaks,\cite{LCB11a,LCB11b} including one that corresponds to a period of $\sim$5--7 unit cells along the (001) direction.  This happens to be in agreement with the oscillation period that we observe (Fig. \ref{fig:fig_theta_vs_n}).  A full theoretical description of helimagnetism in the LNO layer in our samples would need to include the localized Ni$^{2+}$ spins near the interface, magnetic instabilities in the underlying electronic structure, and the interfacial Ni-Mn exchange interaction in a self-consistent manner,\cite{FWS13} which is beyond the scope of this work.

We now construct an effective energy function for the spin-helix model, supposing that the amplitude of the magnetization in the manganite and nickelate layers are $S_{\mathrm{Mn}}$ and $S_{\mathrm{Ni}}$, and that the helix rotates by angle $\theta_0$ from one manganite layer to the neighboring one, as shown schematically in Fig. \ref{fig:fig_pnr}(d).  If we assume that intra-helix stiffness is significantly higher than the coupling to the manganite layers, the interfacial exchange energy in zero magnetic field is
\begin{equation}
F_{\mathrm{S-H}}(\theta) = -2 J S_{\mathrm{Mn}} S_{\mathrm{Ni}} \cos\left(\frac{\theta-\theta_0}{2}\right),
\label{eqn:eqn_FSS}
\end{equation}
where $J$ is the exchange coupling between LSMO and LNO.  In a magnetic field we need to include the Zeeman energy of both the Mn and Ni atoms.  The former is simply $-g \mu_\mathrm{B} S_{\mathrm{Mn}} t H \cos(\theta/2)$, with $t$ the thickness of the LSMO in unit cells, while the latter has to be summed over all of the Ni atoms in the helix.  For a rigid Ni spin helix, we can ignore the Zeeman contribution from the Ni spins, and the magnetic field dependence of the coupling angle may be determined exactly through minimization of Eqn. \ref{eqn:eqn_FSS}, and is given by
\begin{equation}
\theta\left(H\right) = 2 \tan^{-1}\left[\frac{\sin\left(\frac{\theta_0}{2}\right)}{\cos\left(\frac{\theta_0}{2}\right)+\frac{g \mu_\mathrm{B} t H}{2 J S_\mathrm{Ni}}}\right].
\label{eqn:eqn_Theta_SH}
\end{equation}

\begin{figure*}[htbp]
	\includegraphics[width=\textwidth]{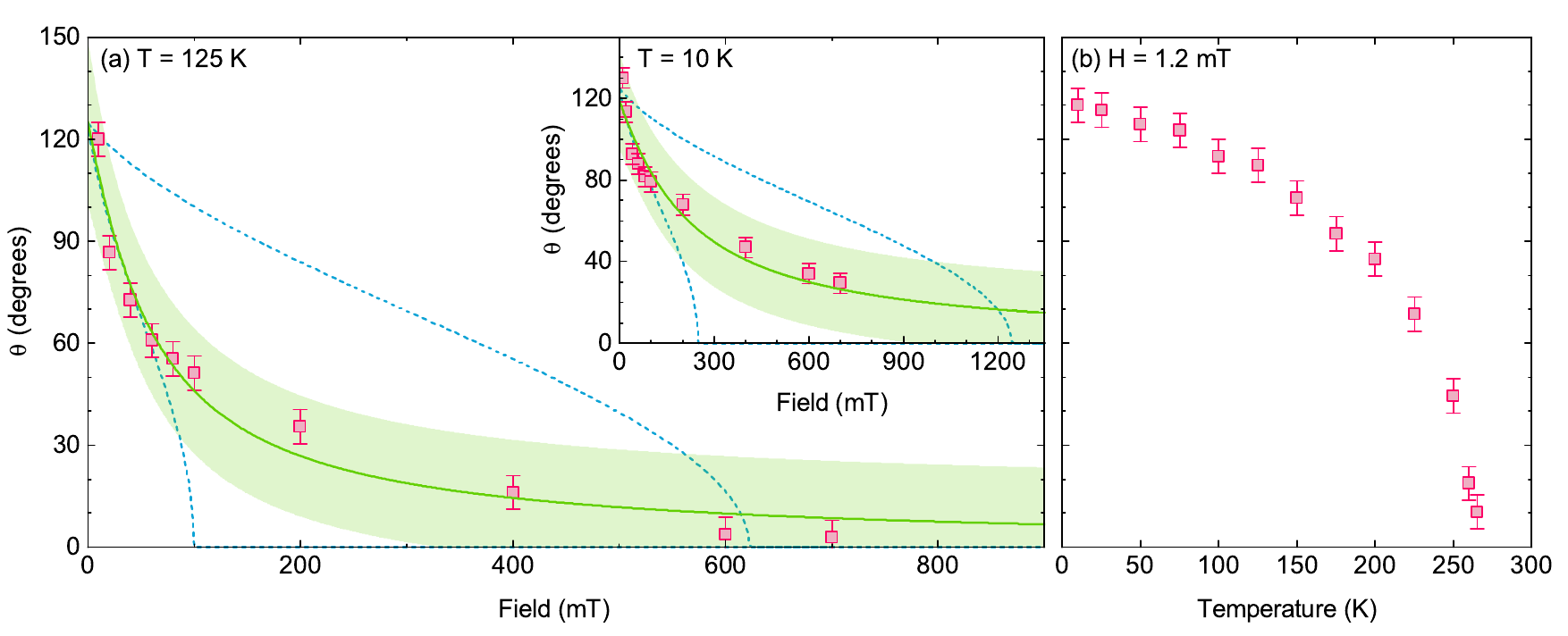}
	\caption{\textbf{Magnetic field and temperature dependence of the coupling angle for [(LaNiO$_3$)$_3$/(La$_{2/3}$Sr$_{1/3}$MnO$_3$)$_9$]$_{14}$ superlattice on LSAT.}  \textbf{(a)} Evolution of the coupling angle with in-plane field applied along the (100) direction at 125 K (a) and 10 K (inset). Solid line shows the best fit to Eqn. \ref{eqn:eqn_Theta_SH} with $J = 3.4 \times 10^{-5}$ J m$^{-2}$ at 125 K and $J = 1.2 \times 10^{-4}$ J m$^{-2}$ at 10 K, assuming $S_\mathrm{Ni}$ = 1 at both temperatures.  Shaded regions represent the 95\% prediction interval using Eqn. \ref{eqn:eqn_Theta_SH}.  Dashed lines show the expected behavior within the bilinear/biquadratic model (Eqn. \ref{eqn:eqn_FBLBQ}) for $J_{\mathrm{BL}} = -1.00 \times 10^{-4}$ J m$^{-2}$ and $J_{\mathrm{BQ}} = -8.75 \times 10^{-5}$ J m$^{-2}$ (upper curve, main panel), $J_{\mathrm{BL}} = -1.60 \times 10^{-5}$ J m$^{-2}$ and $J_{\mathrm{BQ}} = -1.40 \times 10^{-5}$ J m$^{-2}$ (lower curve, main panel), $J_{\mathrm{BL}} = -2.00 \times 10^{-4}$ J m$^{-2}$ and $J_{\mathrm{BQ}} = -1.75 \times 10^{-4}$ J m$^{-2}$ (upper curve, inset), and $J_{\mathrm{BL}} = -4.00 \times 10^{-5}$ J m$^{-2}$ and $J_{\mathrm{BQ}} = -3.00 \times 10^{-5}$ J m$^{-2}$ (lower curve, inset).  \textbf{(b)} Variation of the coupling angle with temperature in an applied field of 1.2 mT.  Error bars $(\pm 5^{\circ})$ indicate the estimated uncertainty in fitting of the PNR spectra, and are greater than the errors that arise from counting statistics.}
	\label{fig:fig_Hdep}
\end{figure*}

To distinguish between the bilinear/biquadratic and spin-helix models, we measure the coupling angle as a function of in-plane magnetic field and temperatures as low as 10 K.  The results for an $n$ = 3 superlattice grown on LSAT and measured at $T$ = 125 K are shown in Fig. \ref{fig:fig_Hdep}(a).  For this sample, the magnetization of neighboring LSMO layers approaches ferromagnetic alignment at around 600 mT, above which no (00$\frac{1}{2}$) spin flip peak is measurable.  We carry out least-squares fitting to Eqn. \ref{eqn:eqn_Theta_SH}, and the resulting fit is shown by the solid line in Fig. \ref{fig:fig_Hdep}(a).  Taking $S_\mathrm{Ni} = 1$, we obtain an exchange coupling between Mn and Ni of $J = 3.4 \times 10^{-5}$ J m$^{-2}$ ($J = 32$ $\mu$eV per interface unit cell), much less than the value predicted for $J_\mathrm{Mn-Ni}$ at the (001) LaNiO$_3$/LaMnO$_3$ interface.\cite{LH13}  In exchange bias systems, comparable differences between the predicted and measured interfacial exchange coupling energy are observed, and their origin remains an open question.\cite{NS99}  Despite the simplicity of this model, it captures two important characteristics of the data: \emph{i}) the sharp initial drop in coupling angle at low fields, and \emph{ii}) the asymptotic approach to alignment at high field.\footnote{For the $n = 3$ superlattice on LSAT, the magnetization of the LSMO layers is not symmetric about the nominal field direction at the lowest measurement fields, unlike the samples prepared on SrTiO$_3$.  This may result from differences in epitaxial strain imposed by the two substrates.  Despite this behavior at low fields, we still find the angle between the magnetization of adjacent LSMO layers is well described by Eqn. \ref{eqn:eqn_Theta_SH}, which is also in agreement with the field-dependence measured on a second sample on SrTiO$_3$.}  

On the other hand, the response calculated within the bilinear/biquadratic model (dashed lines) according to Eqn. \ref{eqn:eqn_FBLBQ} does not agree with the measured field dependence.  Here, we have fixed the ratio $J_{\mathrm{BL}}$/$J_{\mathrm{BQ}}$ to agree with the measured value of $\theta$ at $H < 1.2$ mT, while the magnitude of $J_{\mathrm{BL}}$ and $J_{\mathrm{BQ}}$ were chosen to match either the low-field behavior, or the ferromagnetic alignment field.  While the selected values of $J_{\mathrm{BL}}$ and $J_{\mathrm{BQ}}$ are comparable to those previously reported in a number of conventional metallic\cite{RSH91, SAZ95a, FB96} and oxide\cite{PLC09} systems, we find they are unable to reproduce the observed field-dependence of the coupling angle.  For example, to reproduce the measured alignment field of $\sim$600 mT requires values of $J_{\mathrm{BL}}$ and $J_{\mathrm{BQ}}$ that lead to large discrepancies between predicted and observed behavior at lower fields.  Furthermore, choosing $J_{\mathrm{BL}}$ and $J_{\mathrm{BQ}}$ to give the correct low-field behavior results in a much lower alignment field of only 125 mT.  Additional measurements carried out on the same sample at $T$ = 10 K (Fig. \ref{fig:fig_Hdep}(inset)), demonstrate that the non-collinearity persists to above 700 mT at lower temperatures, the largest field that we are able to apply within the PNR apparatus.

The temperature dependence of the exchange coupling is also unlike that observed in conventional metallic heterostructures, where non-collinear behavior is often restricted to a limited range of temperatures.\cite{FB96, HSC03}  For the $n$ = 3 superlattice on LSAT, we find that the non-collinearity persists to $\sim$265 K (Fig. \ref{fig:fig_Hdep}(b)), close to the ferromagnetic ordering temperature of LSMO in this sample.  This is above the highest reported magnetic ordering temperature for the rare-earth nickelates of $\sim$250 K for Nd$_{1-x}$Sm$_x$NiO$_3$.\cite{TLN92}  We note that interfacial electric fields that lead to a Dzyaloshinsky-Moriya (D-M) interaction with the D-M vector perpendicular to the interface do not affect the chirality of the spin-helix in LNO.\cite{TS10}  However, cooling through $T_C$ in a magnetic field, as we do here, tends to preferentially align the magnetization of the LSMO layers with the applied field, leading to a natural alternation of helicity in the nickelate layers (Fig. \ref{fig:fig_pnr}(d)).

In conclusion, we have demonstrated that interfacial charge transfer drives an oscillatory non-collinear magnetic coupling between ferromagnetic La$_{2/3}$Sr$_{1/3}$MnO$_3$ layers separated by thin LaNiO$_3$ spacers.  The measured field and temperature dependence of the non-collinearity prove that a new mechanism, which can not be explained by a combination of conventional bilinear and biquadratic interactions, is responsible for the interlayer exchange coupling in this system.  To explain the observed behavior, we propose the formation of a helical magnetic state within LNO.  Such a helimagnetic state may arise from the cooperative interaction between localized Ni$^{2+}$ states that result from interfacial charge transfer, and a magnetic instability that is ubiquitous in the rare-earth nickelates.  This mechanism does not require strong spin-orbit coupling or D-M interactions.  Charge transfer and interfacial electronic reconstructions play a critical role in the creation of novel collective phases not only at interfaces between insulating materials, but, as we show here, metallic oxide interfaces as well.  This phenomenon is expected to be broadly applicable to metallic oxides that are derived from Mott insulators.  Furthermore, we envisage applications where electric fields may be used to control the charge transfer, and thus the electronic and magnetic structure, near interfaces between correlated metallic oxides.

\section{Acknowledgments}
We are grateful to Y. H. Liu, M. D. Stiles, J. A. Borchers, and B. B. Maranville for valuable discussions.  Work at Argonne National Laboratory, including the use of the Center for Nanoscale Materials and Advanced Photon Source, was supported by the U.S. Department of Energy, Office of Basic Energy Sciences under contract number DE-AC02-06CH11357. J.D.H, O.G.H., I.M., S.G.E.t.V, and A.B. acknowledge support from Department of Energy, Office of Basic Energy Science, Materials Science and Engineering Division.  We acknowledge the support of the National Institute of Standards and Technology, U.S. Department of Commerce, in providing the neutron research facilities used in this work.  J.K. and J.M.Z. are supported as part of the Center for Emergent Superconductivity, an Energy Frontier Research Center funded by the US Department of Energy, Office of Science, Office of Basic Energy Sciences, under award number DE-AC02-98CH10886.  The Diamond Light Source is acknowledged for beam time allocated on I10 under Proposal Reference No. SI-9626. 

\bibliography{../../library}

\end{document}